\documentclass[12pt]{article}
\usepackage{epsfig,amssymb}
\usepackage{latexsym}

\newcommand{\bq}{\begin{equation}}
\newcommand{\eq}{\end{equation}}
\newcommand{\bqa}{\begin{eqnarray}}
\newcommand{\eqa}{\end{eqnarray}}
\newcommand{\ben}{\begin{enumerate}}
\newcommand{\een}{\end{enumerate}}
\newcommand{\bc}{\begin{center}}
\newcommand{\ec}{\end{center}}
\newcommand{\bqb}{\begin{eqnarray*}}
\newcommand{\eqb}{\end{eqnarray*}}

\def\pr#1#2#3{Phys. Rev. ${\bf{#1}}$, #2 (#3)}
\def\prl#1#2#3{Phys. Rev. Lett. ${\bf{#1}}$, #2 (#3)}
\def\pl#1#2#3{Phys. Lett. ${\bf{#1}}$, #2 (#3)}
\def\prep#1#2#3{Phys. Rept. ${\bf{#1}}$, #2 (#3)}

\def\jhep#1#2#3{JHEP ${\bf{#1}}$, #2 (#3)}
\def\epj#1#2#3{Eur. Phys. J. ${\bf{#1}}$, #2 (#3)}
\def\ijmp#1#2#3{Int. J. Mod. Phys. ${\bf{#1}}$, #2 (#3)}

\def\polon#1#2#3{Acta Phys. Polon. ${\bf{#1}}$, #2 (#3)}

\begin{document}
\pagenumbering{arabic}
\thispagestyle{empty}
\def\thefootnote{\fnsymbol{footnote}}
\setcounter{footnote}{1}

\begin{flushright}
October 14, 2015\\
 \end{flushright}

\vspace{2cm}

\begin{center}
{\Large {\bf Tests of Higgs boson compositeness through the $HHH$ form factor}}.\\
 \vspace{1cm}
{\large G.J. Gounaris$^a$ and F.M. Renard$^b$}\\
\vspace{0.2cm}
$^a$Department of Theoretical Physics, Aristotle
University of Thessaloniki,\\
Gr-54124, Thessaloniki, Greece.\\
\vspace{0.2cm}
$^b$Laboratoire Univers et Particules de Montpellier,
UMR 5299\\
Universit\'{e} Montpellier II, Place Eug\`{e}ne Bataillon CC072\\
 F-34095 Montpellier Cedex 5.\\
\end{center}

\vspace*{1.cm}
\begin{center}
{\bf Abstract}
\end{center}

We show how the $q^2$-dependence of the triple Higgs boson HHH form factor can reveal
the presence of various types of new physics contributions, like new particles
coupled to the Higgs boson or  Higgs boson constituents, without directly observing them.
We compare the effect of such new contributions to the one of higher order
SM corrections to the point-like HHH coupling, due to triangle, 4-leg and s.e. diagrams.
We establish simple analytic expressions describing accurately at high energy
these SM corrections, as well as the examples of new physics contributions.\\

\vspace{0.5cm}

\def\thefootnote{\arabic{footnote}}
\setcounter{footnote}{0}
\clearpage

\section{Introduction}

In spite of the discovery \cite {Higgsdiscov} of the Higgs boson \cite {Higgs},
as expected in the standard model (SM) \cite{Higgssearch},
the SM cannot be the last word, and physics beyond SM (BSM) should exist \cite{BSMexp, BSMth}.

Various types of proposals have been made for this, in particular
Higgs boson compositeness \cite{Hcomp}, leading for example
to anomalous Higgs boson couplings \cite{Anom}, or to couplings of the Higgs boson
to new visible or invisible particles \cite{Portal}.

There are many processes (involving Higgs boson production or decay)
where effective Higgs boson couplings differing from SM predictions
could be observed. However in most of them the Higgs boson
is on-shell and such a  departure could not obviously tell whether it
is caused by Higgs  compositeness.
Also if the Higgs boson is coupled to an invisible
sector, this fact will not be directly observable.

We recall  that the observation  of a suitable Higgs boson
form factor could give answers to such questions.
Indeed a composite particle (like the proton or the pion) should
have a  form factor. This means an appropriate  $q^2$ dependence
(with some scale) in this form factor.
But contrarily to the case of the proton and  the pion, we do not have
a $\gamma HH$ or a $ZHH$ vertex for studying the $H$ form factor.

In this paper we propose to look at the $HHH$ form factor,
that is at a $q^2\equiv s$ dependence of the $HHH$ vertex when one $H$,
with momentum $q$, is off-shell. There are various processes which
involve the $HHH$ coupling, but not necessarily the form factor
with one $H$ far off-shell, like the one we are interested in.
Examples of adequate processes or subprocesses are  $\mu^+\mu^-\to H  \to HH$,
$ZZ \to H  \to HH$, $W^+W^- \to H  \to HH$ and
 $g g \to H \to HH$ (with $g$ denoting the gluon) and  $\gamma\gamma\to H  \to HH$.
The first three of these examples are particularly simple; they involve the usual
initial tree level $H$ coupling and the final $HHH$ form factor.
The last two ones involve a more complex initial  1-loop $H$ coupling.

In  SM, in addition to the point-like $HHH$ coupling, the $HHH$ vertex
receives order $\alpha$ contributions from the usual scalars, fermions
and gauge bosons, due to triangle, bubble with 4-leg couplings
and $H$ self-energy diagrams. We compute them in Sec. 2 and
we illustrate their corresponding (modest) $s$ dependence.
We establish, for each of them, simple analytic expressions valid at high
$s$.

In Sec. 3 we compute examples of new contributions which
could be induced  either by Higgs boson compositeness or by
the couplings of the Higgs boson to a new set of particles.
We also give their corresponding simple high $s$ expressions.
Illustrations show how such contributions can generate spectacular
differences in the $s$ dependence of the $HHH$ form factor,
with respect to the one predicted by the SM.

Conclusions and outlooks are given in Sec. 4.\\

\section{SM contributions to the   $HHH$ form factor.}

The SM prediction for the $HHH$ form factor  consists in a zero order contribution
given by the  point-like coupling
\bq
eg_{HHH}=-~{3em^2_H\over 2s_Wm_W}~~, \label{HHH-coupling}
\eq
and the higher order   corrections.
We compute them when one $H$ is off-shell, with its squared four-momentum denoted
as $q^2\equiv s$, while the two other $H$ (with four-momenta $p,p'$) are on-shell.

At first $\alpha$ order,
these corrections  consist in
1-loop triangles, bubbles with a 4-leg couplings and $H$ self-energy.
Up to this  order, this form factor is written as
\bq
F^{SM}(s)=eg_{HHH}+A^{SM}(s) ~~, \label{HHH-ASM}
\eq
where in $A^{SM}(s)$ the aforementioned 1-loop SM corrections are collected.

Before discussing them, we  mention that these 1-loop diagrams contain divergent
terms which should be canceled by appropriate counter terms.
There are various possibilities for making these  cancelations,
which differ by small constant terms. As we are essentially interested
only in the high $s$ dependence, we will use the supersimple renormalization scheme
(SRS) procedure \cite{SRS, WW}, which
leads to the simplest expressions in terms of augmented Sudakov logarithms.
Among them,  we will only need (see \cite{SRS,WW} for details):
\bqa
&& \overline{\ln^2s_X}\equiv
\ln^2 s_X +4L_{HXX} ~~, ~~~
s_X \equiv \left (\frac{-s-i\epsilon}{m_X^2} \right ) ~~, \label{Sud-ln2} \\
&& \overline{\ln s_{ij}}\equiv \ln s_{ij}+b^{ij}_0(m_H^2)-2 ~~,~~~
\ln s_{ij}\equiv \ln \frac{-s-i\epsilon}{m_im_j} ~~, \label{Sud-ln}
\eqa
 where explicit expressions for $b_0^{ij}(m^2_H)$ and $L_{HXX}$
 are given e.g. in  Eqs.(A.6, A.5) of \cite{WW}.

 We note that the counter terms needed
 in the SRS scheme  respect this augmented Sudakov structure \cite{SRS,WW}.
Globally this procedure consists in replacing the divergent terms
related to the  $(i,j)$ internal lines of any contributing   diagram,   as
\bq
\ln \frac{-s-i\epsilon}{\mu^2} -\Delta \to \ln s_{ij}+b_0^{ij}(m^2_H)
~~, \label{SRS-replacement}
\eq
where $\mu$ denotes the renormalization scale and $\Delta=1/\epsilon -\gamma_E +\ln (4\pi)$,
with  the number of dimensions used for regularization written as $n=4-2\epsilon$.

In the present case, with only  triangle and bubble diagrams contributing,
there is no ambiguity related to the internal lines $(i,j)$.
They can only be  $H$, $Z$, $W$ and $t$, so that we can only have $(ij)=(HH),(ZZ),(WW), (tt)$.
The SRS results thus obtained, are always denoted as ``sim"
in the illustrations.\\

We next describe  the exact   expressions for the various
triangle and bubble diagrams with 4-leg couplings,  as well as
 their high energy SRS (sim) forms.\\

\subsection{Triangles and bubbles with 4-leg couplings}

\noindent
{\bf Scalar $(SSS)$ triangles  and $(SS)$ bubbles with a 4-leg $SSHH$ coupling.}
\bqa
&&A^{SM}_{SSS}(s)=-~{e\alpha\over4\pi}\Big \{g^3_{HSS}
C_0(s,m^2_H,m^2_H,m^2_S,m^2_S,m^2_S) \nonumber\\
&& +g_{HSS}g_{HHSS}[B_0(s,m^2_S,m^2_S)+2B_0(m^2_H,m^2_S,m^2_S] \Big \}\nonumber\\
&& \to -~{e\alpha\over4\pi}\Big  \{g^3_{HSS}{ \overline{ \ln^2s_S}\over s}
+g_{HSS}g_{HHSS}(-  \overline{\ln s_{SS}}) \Big \}~~. \label{ASMSSS}
\eqa
This applies to $SSS = HHH, G^0G^0G^0, C^ZC^ZC^Z, G^{\pm}G^{\pm}G^{\pm},
C^{\pm}C^{\pm}C^{\pm}$ triangles and to $SS = HH, G^0G^0, G^{\pm}G^{\pm}$ bubbles,
where $g_{HHH}$  is given in (\ref{HHH-coupling})
and\footnote{$G^\pm, G^0$ denote the SM Goldstone fields and $C^\pm, C^Z$
FP ghosts.}
\bqa
&& g_{HHHH}=-~{3m^2_H\over 4s^2_Wm^2_W} ~,~
g_{HGG}=-~{m^2_H\over 2s_Wm_W}  ~,~  g_{HHGG}={1\over 2s^2_Wc^2_W} ~, \nonumber \\
&& g_{HC^ZC^Z}=-~{m_W\over 2s_Wc^2_W} ~,~ g_{HC^{\pm}C^{\pm}}=-~{m_W\over 2s_W} ~.
\label{couplings1}
\eqa
Note that for the ghost loop, there is no 4-leg diagram, and that
 a global fermionic minus sign has been inserted.
In all cases, the internal $S$ mass
for $H, G^0, C^Z, G^\pm, C^\pm$ is respectively equal to the one of $H,Z,Z,W, W$.

\vspace{1.cm}
\noindent
{\bf Fermion triangles $(fff)$}\\
Due to the strong mass dependence of the $Hff$ coupling,
it is adequate to restrict to the $(ttt)+(\bar t\bar t\bar t)$ case. The result is
\bqa
&&A^{SM}_{ttt}(s)=~{e\alpha\over4\pi} {3m^3_t\over2s^3_Wm^3_W}
\Big \{2m^3_tC_0+2m_t[3m^2_H(C_{21}+C_{22})+6p.p'C_{23}\nonumber\\
&& +3nC_{24} +2q.pC_{11}+2q.p'C_{12}+2m^2_HC_{11}+2p.p'C_{12}+q.pC_{0}]\Big \}
\nonumber\\
&& \to ~{e\alpha\over4\pi} {3m^3_t\over2s^3_Wm^3_W}
\Big \{2m_t \Big [{- \overline{\ln^2s_t} \over4}- \overline{ \ln s_{tt}} \Big] \Big \}
~~. \label{ASMttt}
\eqa

\vspace{1.cm}
\noindent
{\bf Vector triangles $(VVV)$ and bubbles $(VV)$  with a 4-leg $HHVV$ coupling.}
\bqa
&&A^{SM}_{VVV}(s)={e\alpha\over4\pi} \Big \{g^3_{HVV}nC_0
+g_{HVV}g_{HHVV}[2B_0(m^2_H,m^2_V,m^2_V)+B_0(s,m^2_V,m^2_V)] \Big \}
\nonumber\\
&&\to -~{e\alpha\over4\pi} \Big \{g^3_{HVV}{2\overline{\ln^2s_V}\over s}
+g_{HVV}g_{HHVV}[- \overline{\ln s_{VV}}] \Big \} ~~, \label{ASMVVV}
\eqa
applied only to $V=Z,W$, since there are no $HZ\gamma$ or $HHZ\gamma$ couplings.
Because of this, the $V$ masses in the SRS forms  $\overline{ \ln^2s_V}$ and $\overline{ \ln s_{VV}}$
 can either be   $m_Z$ or  $m_W$.

\vspace{1.cm}
\noindent
{\bf $(VVS)$ triangles}
\bqa
&&A^{SM}_{VVS}(s)={e\alpha\over4\pi}g^2_{VSH}g^2_{VVH} \Big \{
m^2_H(C_{21}+C_{22})+2p.p'C_{23}+nC_{24}
\nonumber\\
&& +(p.p'+3q.p)C_{11}+(m^2_H+3q.p')C_{12}+2(q^2+q.p')C_{0}\Big \}
\nonumber\\
&&\to {e\alpha\over4\pi}g^2_{VSH}g^2_{VVH}\Big \{
{1\over2}( \overline{\ln^2s_V}+ \overline{\ln s_{VV}}) \Big \}~~, \label{ASMVVS}
\eqa
applied to $ZZG^0, W^{\pm}W^{\pm}G^{\pm}$; compare (\ref{ASMVVV}).\\

\vspace{1.cm}
\noindent
{\bf $(VSV)$ triangles}
\bqa
&&A^{SM}_{VSV}(s)={e\alpha\over4\pi}g^2_{VSH}g^2_{VVH}\Big \{
m^2_H(C_{21}+C_{22})+2p.p'C_{23}+nC_{24}
\nonumber\\
&& +(3m^2_H-p.p')(C_{11}-C_{12})+2(m^2_H-p.p')C_{0} \Big \}
\nonumber\\
&&\to {e\alpha\over4\pi}g^2_{VSH}g^2_{VVH} \Big \{
{1\over4} \overline{ \ln^2s_V}+2 \overline{ \ln s_{VV}} \Big \}~~, \label{ASMVSV}
\eqa
applied to $ZG^0Z, W^{\pm}G^{\pm}W^{\pm}$.

\vspace{1.cm}
\noindent
{\bf $(SVV)$ triangles}
\bqa
&&A^{SM}_{SVV}(s)={e\alpha\over4\pi}g^2_{VSH}g^2_{VVH}\Big \{
m^2_H(C_{21}+C_{22})+2p.p'C_{23}+nC_{24}
\nonumber\\
&& -(m^2_H+q.p)C_{11}-(p.p'+q.p')C_{12})+q.pC_{0} \Big \}\nonumber\\
&&\to {e\alpha\over4\pi}g^2_{VSH}g^2_{VVH}\Big \{
-{1\over2}( \overline{ \ln^2s_V}+ \overline{ \ln s_{VV}})\Big \}~~, \label{ASMSVV}
\eqa
applied to $G^0ZZ, G^{\pm} W^{\pm}W^{\pm}$.

\vspace{1.cm}
\noindent
{\bf $(VSS)$ triangles}
\bqa
&&A^{SM}_{VSS}(s)={e\alpha\over4\pi}g^2_{VSH}g^2_{SSH}\Big \{
-m^2_H(C_{21}+C_{22})-2p.p'C_{23}-nC_{24}
\nonumber\\
&& -2(m^2_H+q.p)C_{11}-2(p.p'+q.p')C_{12})-4q.pC_{0}\Big \}
\nonumber\\
&&\to {e\alpha\over4\pi}g^2_{VSH}g^2_{SSH}\Big \{
-{1\over2} \overline{ \ln^2s_V}+ \overline{ \ln s_{VV}}\Big \}~~, \label{ASMVSS}
\eqa
applied to $ZG^0G^0, W^{\pm}G^{\pm} G^{\pm}$.

\vspace{1.cm}
\noindent
{\bf $(SVS)$ triangles}
\bqa
&&A^{SM}_{SVS}(s)={e\alpha\over4\pi}g^2_{VSH}g^2_{SSH}\Big \{
-m^2_H(C_{21}+C_{22})-2p.p'C_{23}-nC_{24}
\nonumber\\
&& -(-m^2_H+q.p+p.p')C_{11}-(m^2_H-p.p'+q.p')C_{12})+(p.p'+q.p)C_{0}\Big \}
\nonumber\\
&&\to {e\alpha\over4\pi}g^2_{VSH}g^2_{SSH}
\{ \overline{ \ln^2s_V}- \overline{ \ln s_{VV}}\}~~, \label{ASMSVS}
\eqa
applied to $G^0ZG^0, G^{\pm}W^{\pm} G^{\pm}$.

\vspace{1.cm}
\noindent
{\bf $(SSV)$ triangles}
\bqa
&&A^{SM}_{SSV}(s)={e\alpha\over4\pi}g^2_{VSH}g^2_{SSH} \Big \{
-m^2_H(C_{21}+C_{22})-2p.p'C_{23}-nC_{24}
\nonumber\\
&& -(m^2_H-q.p-p.p')C_{11}-(-m^2_H+p.p'-q.p')C_{12})+(-q.p'+q.p)C_{0}\Big \}
\nonumber\\
&&\to {e\alpha\over4\pi}g^2_{VSH}g^2_{SSH}\Big  \{
-{1\over2} \overline{ \ln^2s_V}+ \overline{ \ln s_{VV}}\Big  \}~~, \label{ASMSSV}
\eqa
applied to $G^0G^0Z, G^{\pm}G^{\pm}W^{\pm}$.\\

In the above contributions the following couplings are needed
\bqa
&& g_{ZZH}={m_Z\over s_wc_w} ~,~   g_{ZZHH}={1\over 2s^2_wc^2_w} ~,~ \nonumber \\
&& g_{WWH}={m_W\over s_w}  ~,~  g_{WWHH}={1\over 2s^2_w} ~,~
 g_{ZGH}=g_{WGH}={1\over 2s_wc_w} ~~ . \label{couplings2}
\eqa\\

\subsection{$H$ self-energy}

This additional contribution is given by
\bq
A^{SM}_{se}(s)=-~{eg_{HHH}\over s-m^2_H}\Sigma_H(s)~~, \label{ASMse}
\eq
where $\Sigma_H(s)$ is computed from the following diagrams:

\vspace{1.cm}
\noindent
{\bf Bubbles $VV$ } leading to
\bq
\Sigma_H(s)
={X^2_1\over4\pi^2}[B_0]\to
{X^2_1\over4\pi^2}[- \overline{ \ln s_{VV}}] ~~,\label{SigmaHVV}
\eq
for which we respectively get
\bqa
VV=ZZ &\to &  X^2_1={e^2M^2_W\over 2s^2_Wc^4_W} ~~, \nonumber \\
VV=W^{\pm}W^{\mp} &\to &  X^2_1={e^2M^2_W\over s^2_W}~~. \label{VVX2}
\eqa

\vspace{1.cm}
\noindent
{\bf Bubbles $SV$ } leading to
\bqa
\Sigma_H(s)
 && =-~{X^2_1\over16\pi^2}[s(B_0+B_{21}-2B_1)+nB_{22}]
 \nonumber\\
&& \to -~{X^2_1\over16\pi^2}[-2s \overline{\ln s_{SV}}]~~,\label{SigmaHSV}
\eqa
for which we respectively get
\bqa
SV=G^0Z  &\to & X^2_1={e^2\over4s^2_Wc^2_W}~~, \nonumber \\
SV=G^{\mp}W^{\pm} &\to &  X^2_1={e^2\over2s^2_W}~~. \label{SVX2}
\eqa

\vspace{1.cm}
\noindent
{\bf \underline{Bubble $tt$}} leading to
\bqa
&&\Sigma_H(s)
=-~{1\over4\pi^2}[(s(B_1+B_{21})+nB_{22}+m^2_tB_0)X^2_1]\nonumber\\
&&\to
-~{X^2_1\over4\pi^2}[{s\over2} \overline{\ln s_{tt}}]~~,\label{SigmaHtt}
\eqa
with
\bq
X^2_1={3e^2\over 4s^2_WM^2_W}[m^2_t]~~. \label{ttX2}
\eq

\vspace{1.cm}
\noindent
{\bf Bubbles $SS$} leading to
\bq
\Sigma_H(s)
={X^2_1\over16\pi^2}[B_0] \to
{X^2_1\over16\pi^2}[- \overline{\ln s_{SS}}]~~,\label{SigmaHSS}
\eq
with
\bq
X^2_1 = {9e^2m^4_H\over8s^2_WM^2_W}~,~
{e^2m^4_H\over8s^2_WM^2_W}~,~{e^2m^4_H\over4s^2_WM^2_W}~,~
-{e^2m^2_W\over4s^2_Wc^4_W}~,~-{e^2m^2_W\over2s^2_W}~, \label{SSX2}
\eq
for
\bq
SS=~~HH,~~G^0G^0 ~,~ G^+,G^- ~,~ C^ZC^Z~,~ C^+C^-~~, \label{SSX2a}
\eq
respectively. Note that in these $SS$ bubbles, the internal $S$ mass is correspondingly
equal to the mass of $H,Z,W,Z,W$.\\

\subsection{Illustrations}

Summing the above  contributions, either the exact ones
or their high energy SRS ``sim" forms,
we obtain in Figs.1 the SM predictions for $A^{SM}$ as functions of $\sqrt{s}$.

In the upper row of Fig.1, the left panel corresponds to $(ttt)$ obtained from (\ref{ASMttt}),
the middle panel to $(HHH)$ derived from (\ref{ASMSSS}), while the right panel refers
to $(VVV+GGG+CCC)$ derived from (\ref{ASMSSS}, \ref{ASMVVV}).

In the middle row, the left panel corresponds to $(VVS+VSV+SVV)$ obtained from
(\ref{ASMVVS},\ref{ASMVSV},\ref{ASMSVV}), the middle panel to $(VSS+SVS+SSV)$
obtained from (\ref{ASMVSS}, \ref{ASMSVS}, \ref{ASMSSV}), while the right panel shows
 the  Higgs self energy contribution derived from the expressions in Subsection 2.2.

Finally, in the panel at the lowest row, we present the complete $F^{SM}$ result of
(\ref{HHH-ASM}).

As seen from Fig.1, the $ttt$ part (upper left panel)
is quickly dominating   above the tt threshold.

Fig.1 also shows that the shapes of the real and imaginary parts
of the various contributions are quite  different. This may be useful  in case of a
departure from the SM prediction is observed\footnote{See examples in the next Section.}.
It could for example  suggest what type of new contribution should be added.

Note also that the SRS (sim) approximation is globally OK above $\sim 1.5$TeV.\\

\section{Examples of new physics contributions.}

In a pure compositeness picture there is no Born $HHH$ point-like vertex.
The whole $HHH$ coupling should then come from an effective $(XXX)$ triangles made by the
constituents $X$ and an effective $HXX$ coupling related to the binding.

On another hand, if the Higgs boson is connected to a new sector, one may have
triangles involving the corresponding new particles.

In the case of a strong sector (similarly to the hadronic case),
there may be resonances $R$ leading to $HHH$ contributions
of the type $H\to R(XX) \to HH$.

In Figs.2 we have made illustrations of the contributions to the $(HHH)$ form factor
corresponding to these examples.

\begin{itemize}

\item
For a scalar Higgs-constituent $X$, we get
\bqa
&&A_{XXX}(s)=-~{e\alpha\over4\pi}g^3_{HXX}
C_0(s,m^2_H,m^2_H,m^2_X,m^2_X,m^2_X)
\nonumber\\
&& \to -~{e\alpha\over4\pi}g^3_{HXX}{ \overline{ \ln^2s_X}\over 2s} ~~, \label{ANPXXX}
\eqa

\item
for a fermionic constituent $F$, we get
\bqa
&&A_{FFF}(s)=-~{e\alpha\over4\pi}g_{HFF}^3
\Big \{2m^3_FC_0+2m_F \Big [3m^2_H(C_{21}+C_{22})+6p.p'C_{23}+3nC_{24}
\nonumber\\
&& +2q.pC_{11}+2q.p'C_{12}+2m^2_HC_{11}+2p.p'C_{12}+q.pC_{0}\Big ]\Big \}
\nonumber\\
&&\to -~{e\alpha\over4\pi}g_{HFF}^3
\{2m_F[{- \overline{ \ln^2s_F} \over4}- \overline{ \ln s_{FF}}]\}~~, \label{ANPFFF}
\eqa

\item

and for a typical resonance effect  we get
\bq
A_{Res}(s)= {g_{HR}g_{RHH}\over s-M^2_R+iM_R\Gamma_R} ~~. \label{ANPRes}
\eq

\end{itemize}

Effective couplings have been  chosen so that the
resulting $HHH$ form factor is similar to the SM one. In such a case,
the illustrations serve to emphasize the different $s$-dependencies. In Figs.2a,b,c one indeed sees that the $s$ dependencies appearing in these
examples are very different from each other and also very different from the SM
case.

So we believe that there is much to be learned  from the measurement of the $HHH$ form
factor.\\

\section{Conclusions and outlooks.}

We have computed the real and imaginary parts of the  SM 1-loop contributions
to the $HHH$ form factors, as well as examples of possible new physics effects
corresponding either to
Higgs boson compositeness or to the coupling of the Higgs boson to a new sector.

We have emphasized the fact that the $q^2\equiv s$ dependencies
of the $HHH$ form factor can be very different, depending on
their origin. As it can be seen in the illustrations
these differences can be spectacular and reflect the specific nature
of the new physics (bosonic or fermionic constituents or resonances).

In each case we have also given the corresponding analytic expressions
in the adequate ``sim" approximation, allowing a quick estimate of the effect at high $s$.

The aim of this paper was only to to put forward this idea of looking
especially at the $s$-dependence of the form factors  and to show
that such effects may exist.

We hope that these results will encourage further phenomenological
and experimental studies of the possibilities to measure the $HHH$ form
factor.

In the simplest situation, this form factor appears in
the process $\mu^-\mu^+ \to HH$, (about the $\mu^-\mu^+$ collider project
see \cite{mumu}). It appears through the s-channel $H$ exchange diagram, which is proportional
to  the $H\mu\mu$ coupling.
There exist  also $t$ and $u$ channel $\mu^\mp$ exchange diagrams,
 but their contribution is suppressed, since it is quadratic in the
 $H\mu\mu$ coupling. Complete 1-loop corrections to these Born terms are then needed,
 in order to be able to compare the effects of new $HHH$ contributions,
to the exact SM prediction, see \cite{mumuhh}.

The subprocesses $ZZ \to HH$ and $W^+W^- \to HH$ also involve the simple
s-channel $H$ exchange diagram associated with  a 4-leg $ZZHH$ or $W^+W^-HH$
vertex. In addition,   $t$ and $u$ channel $Z$ or $W$ exchanges  contribute.
These subprocesses can be reached
at $e^-e^+$ colliders or at LHC, by making a detailed analysis.

The processes  $gg \to HH$ and $\gamma\gamma \to HH$
contain an s-channel $H$ exchange, but the initial vertex needs
a 1-loop contribution. There are also other 1-loop
diagrams producing the final $HH$ state.
Specific works should be devoted to each of these processes,
see e.g. \cite{Dawson, Telnov, Asakawa, Levy}.\\


\begin{figure}[t]
\[
\epsfig{file=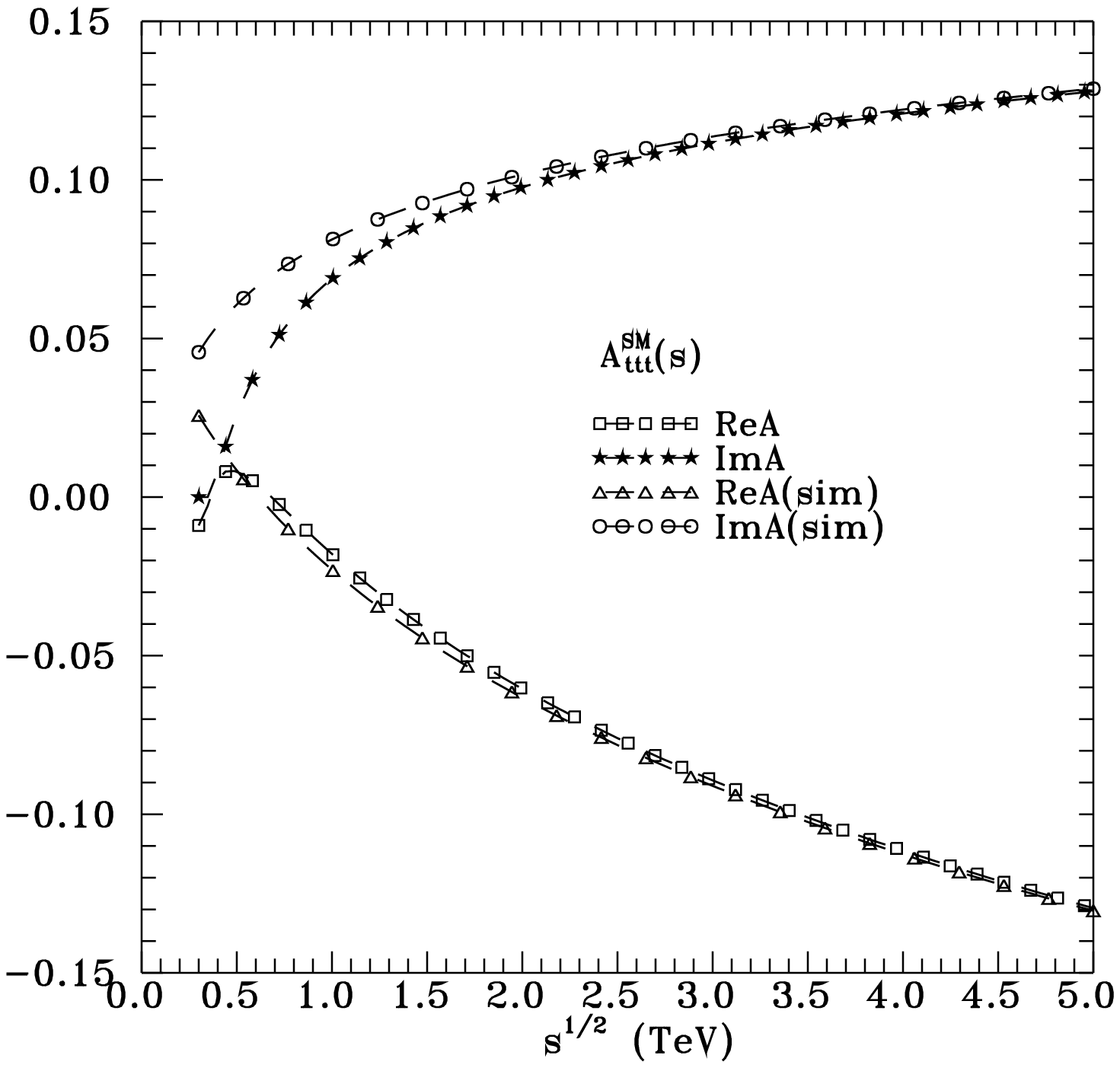, height=4.cm}\hspace{0.5cm}
\epsfig{file=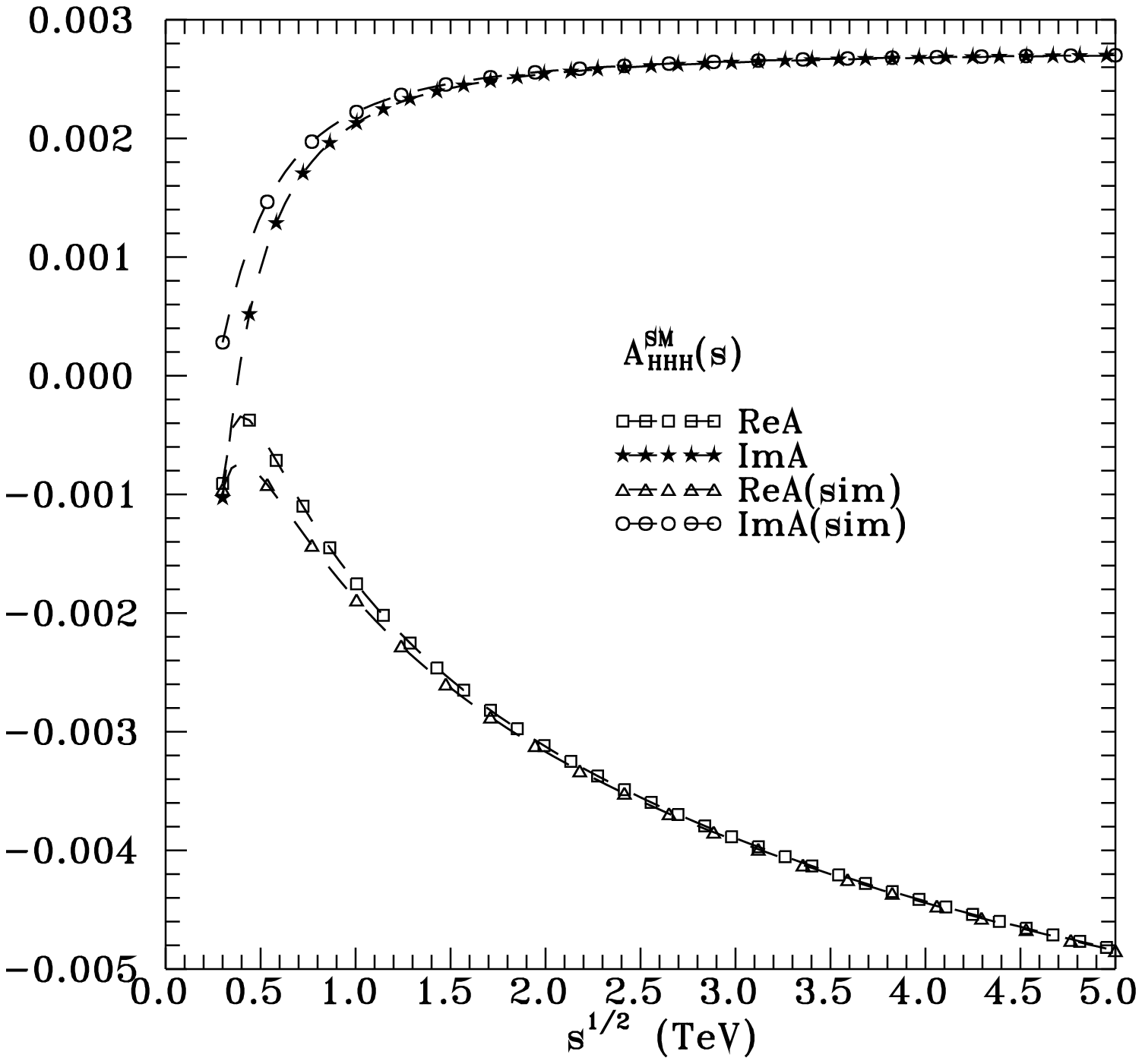,height=4.cm}\hspace{0.5cm}\epsfig{file=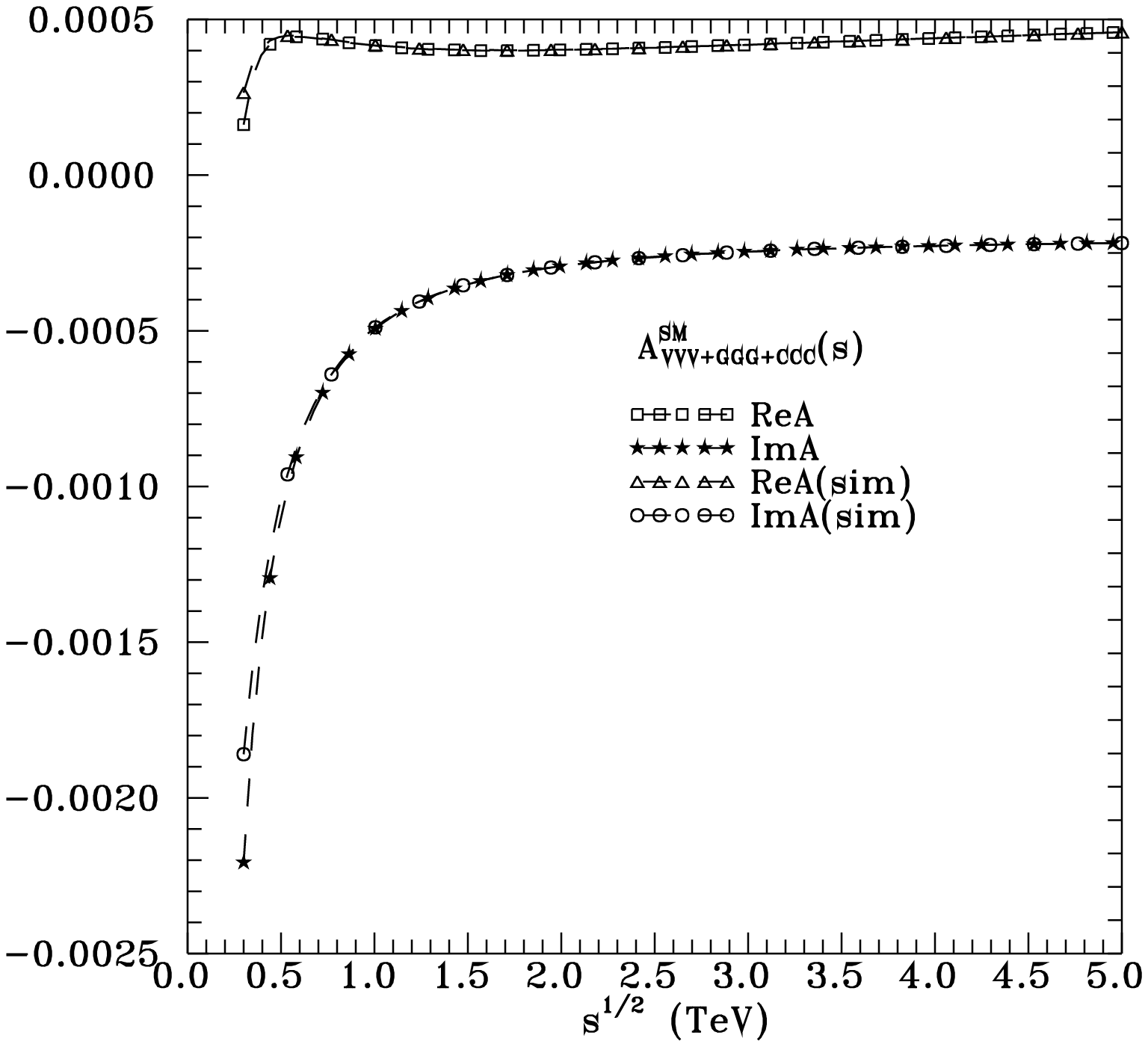, height=4.cm}
\]
\[
\epsfig{file=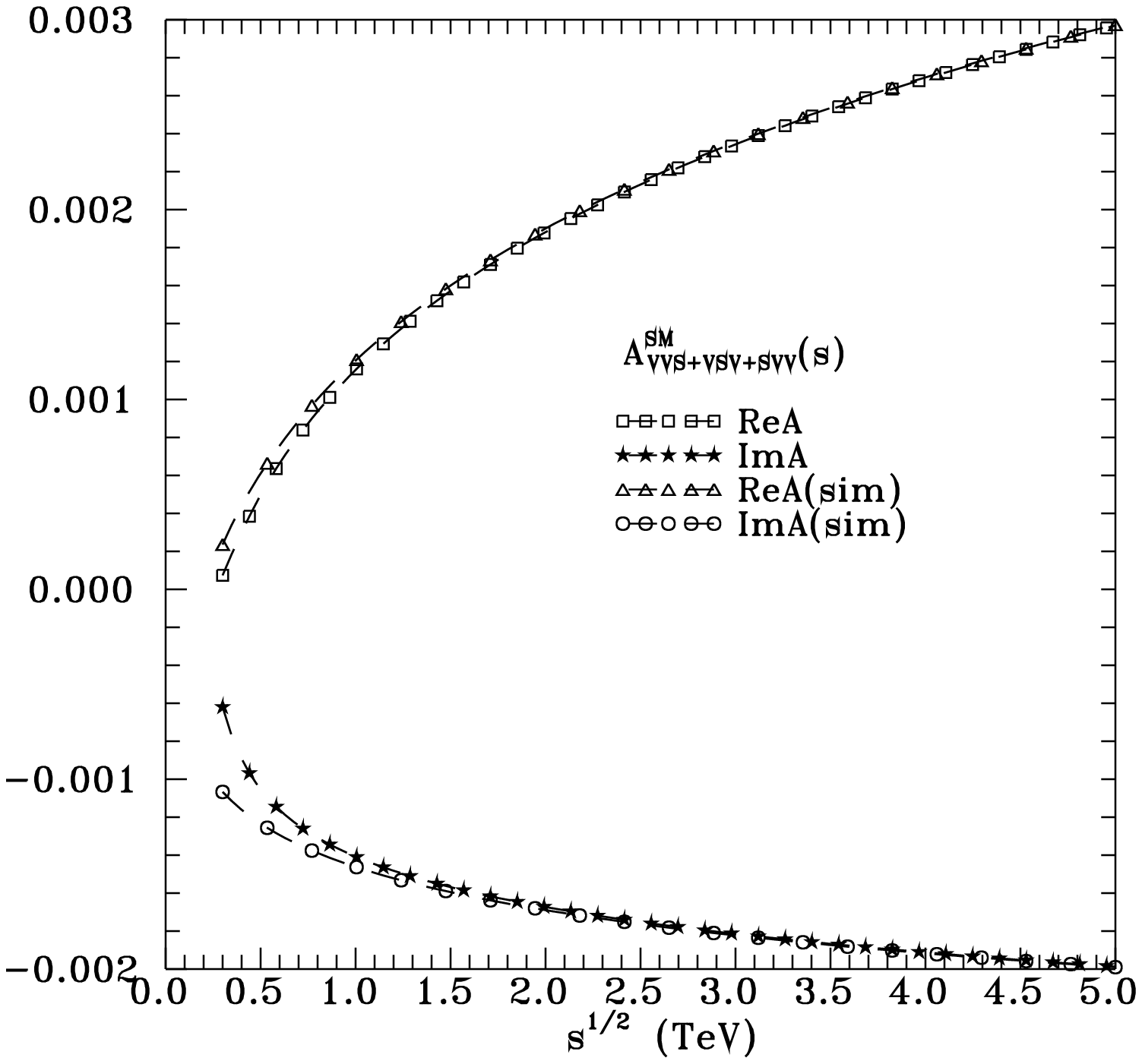, height=4.cm}\hspace{0.5cm}
\epsfig{file=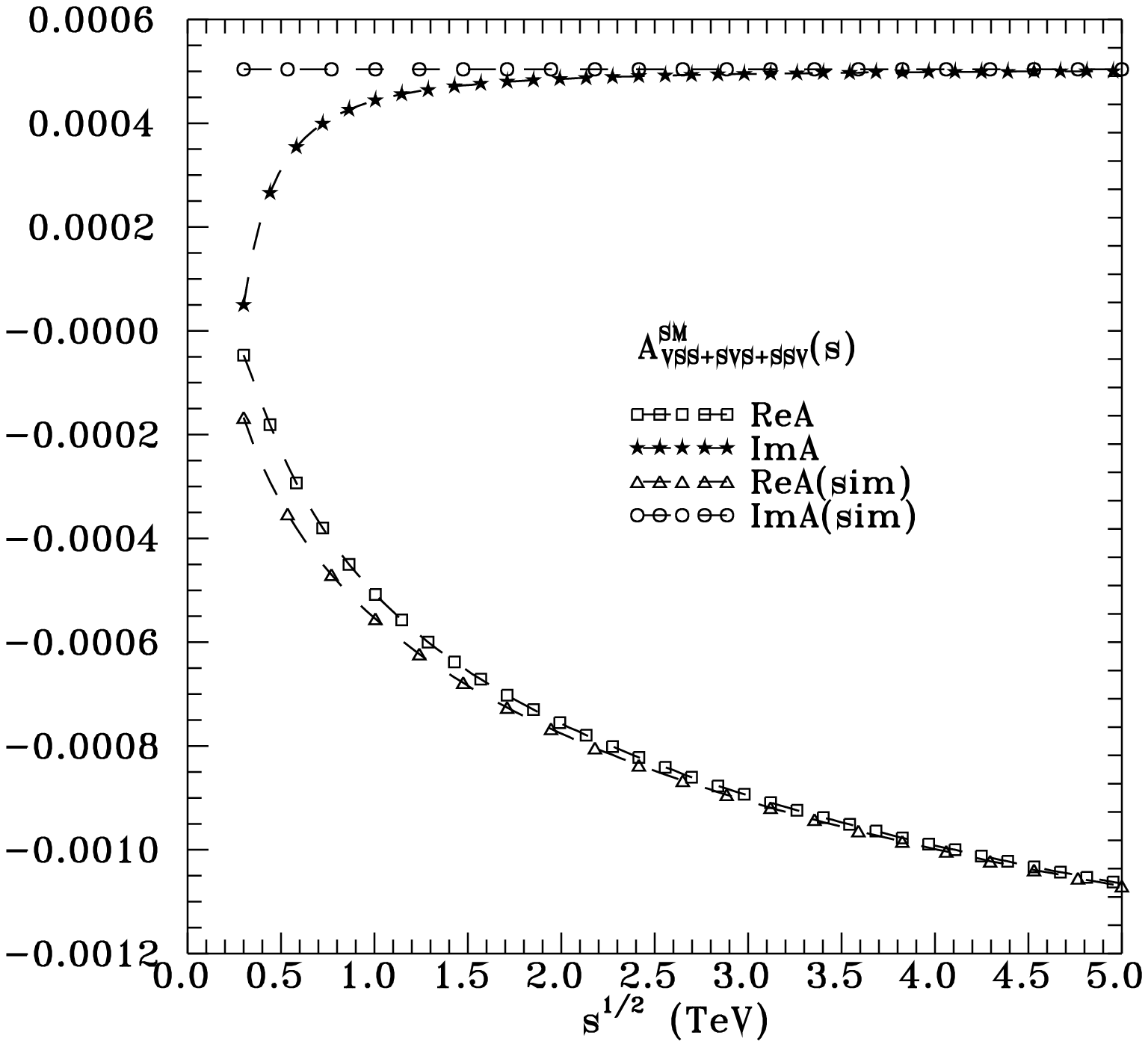,height=4.cm}\hspace{0.5cm}\epsfig{file=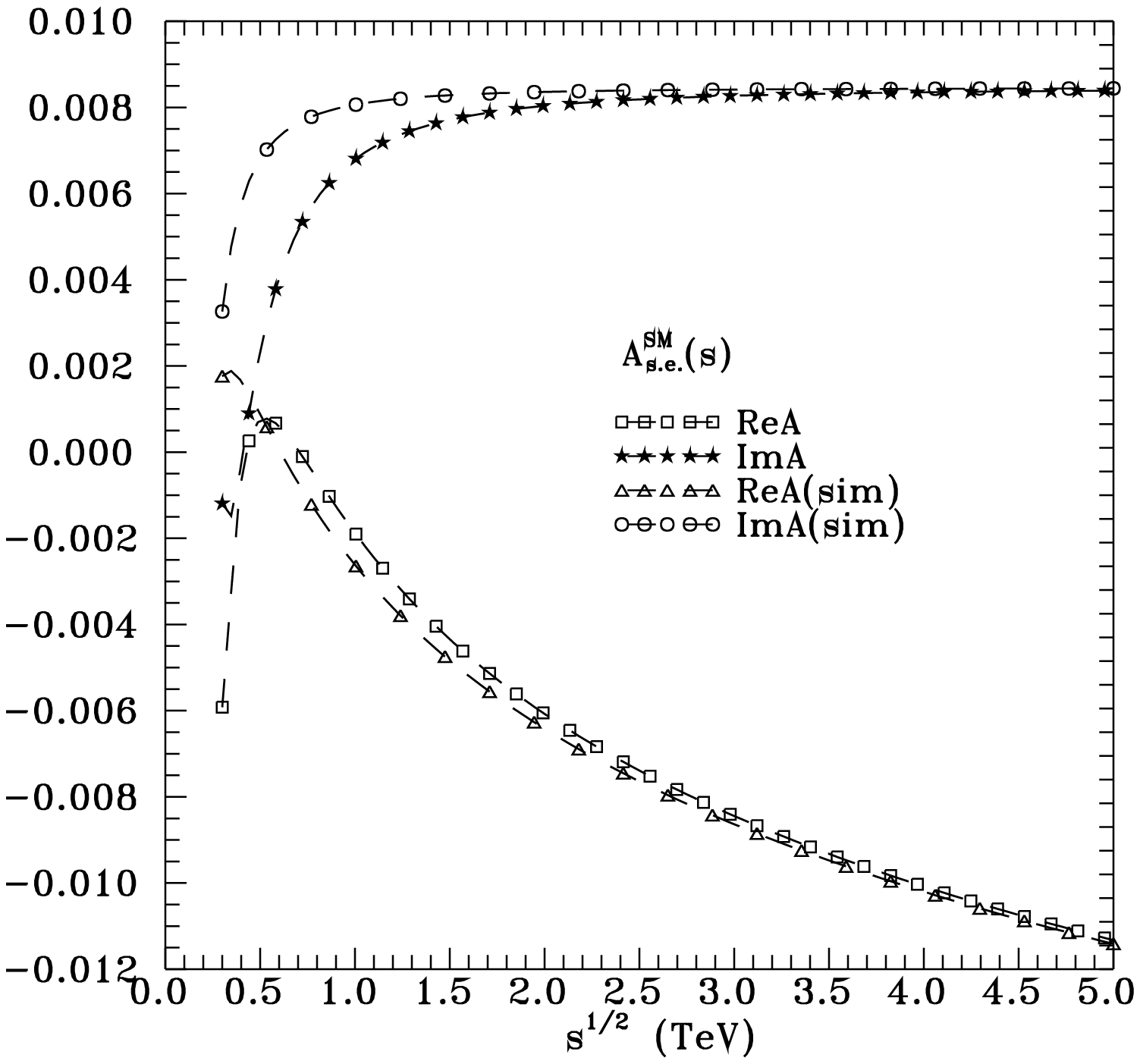,height=4.cm}
\]\vspace{0.1cm}
\[
\epsfig{file=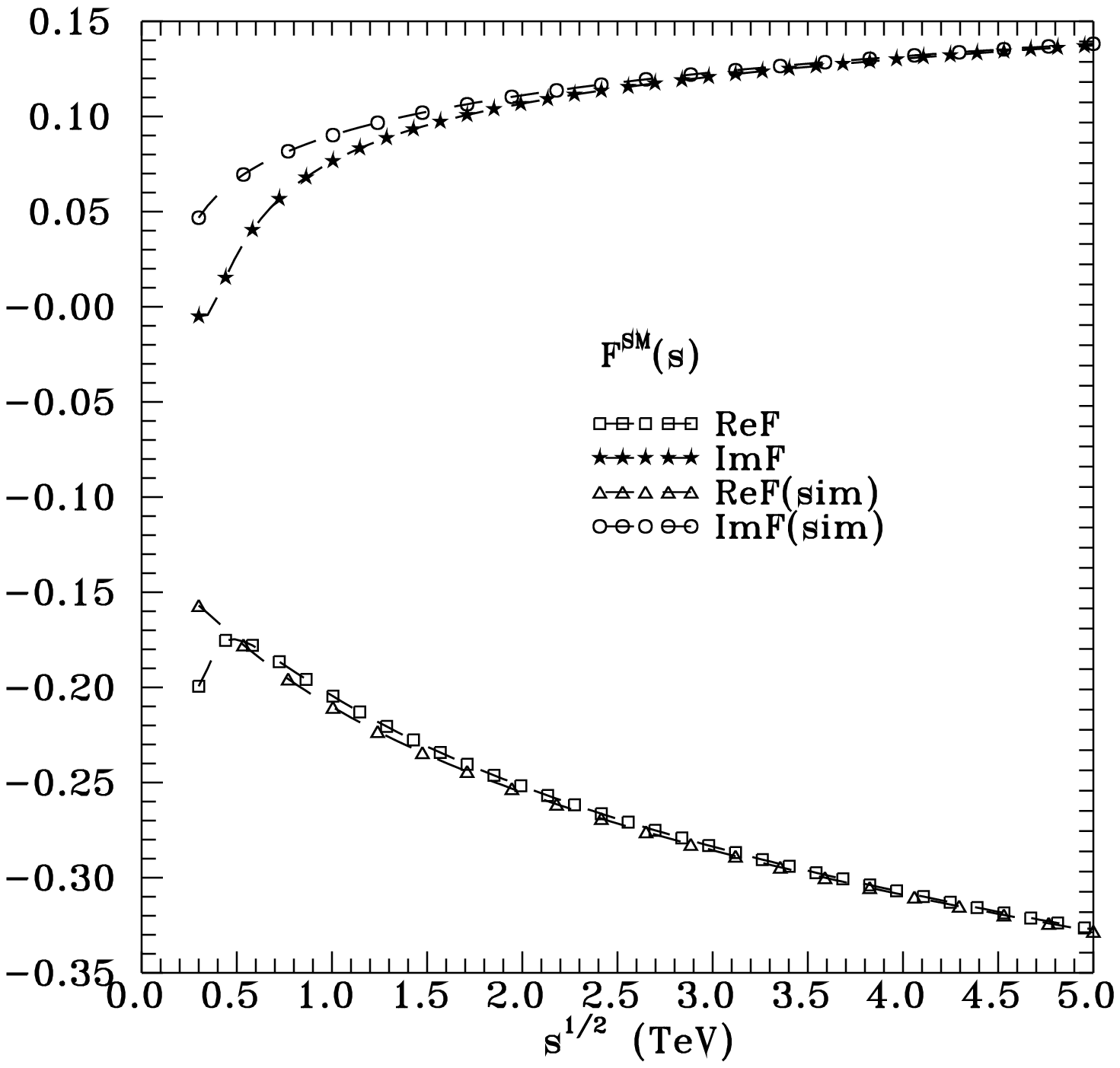, height=7.cm}
\]
\caption[1]{The upper two rows  describe the real and imaginary parts
of the various SM contributions to the  $A^{SM}$ form factors in (\ref{HHH-ASM}),
in the  exact 1-loop treatment and in its high energy
SRS (``sim") approximation; see text in Section 2.3.
The panel in the third row gives the corresponding total SM contribution $F^{SM}$
defined in (\ref{HHH-ASM}).}
\label{Fig1}
\end{figure}

\clearpage

\begin{figure}[h]
\vspace{-1cm}
\[
\epsfig{file=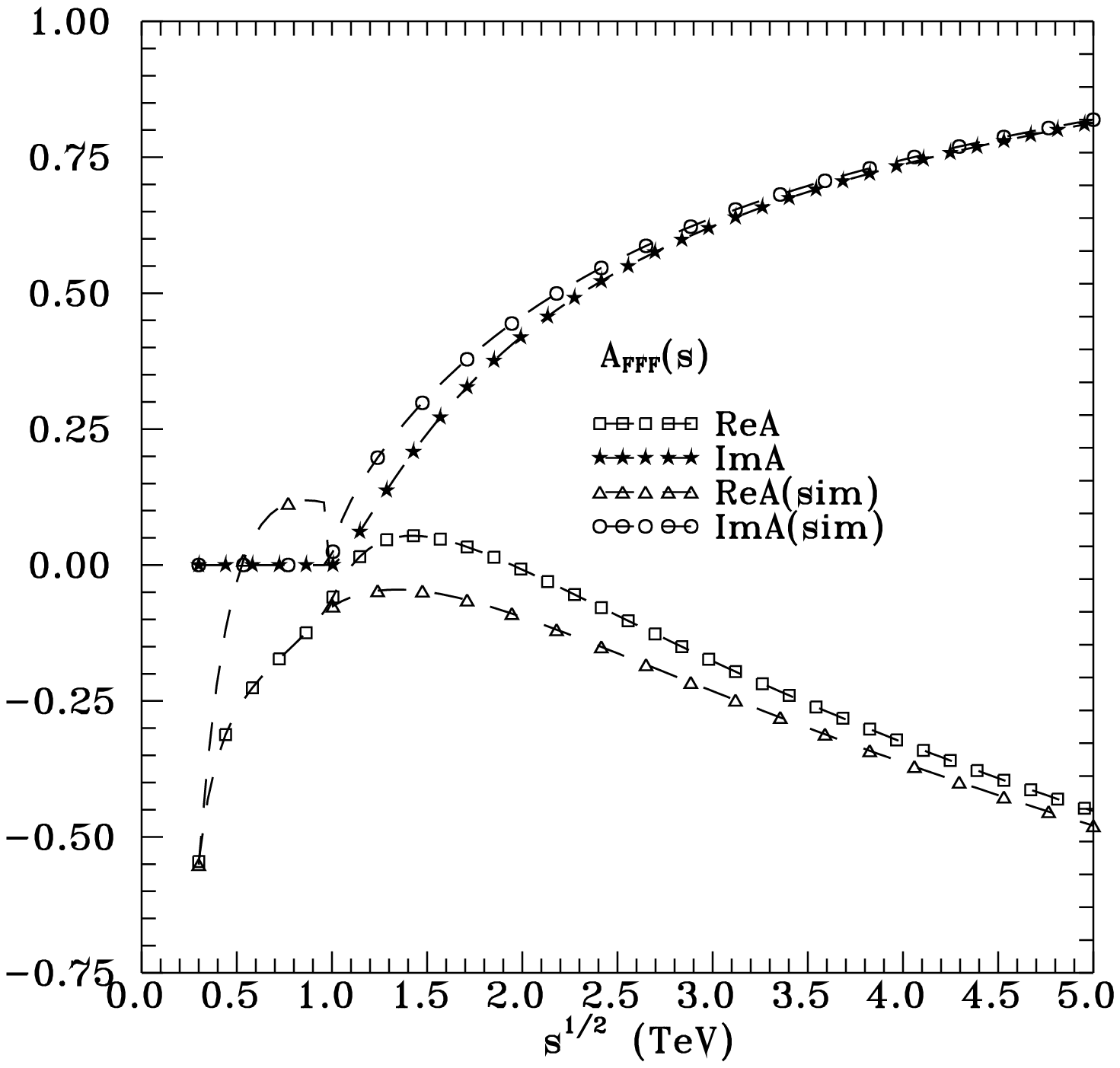, height=7.cm}\hspace{1.0cm}
\epsfig{file=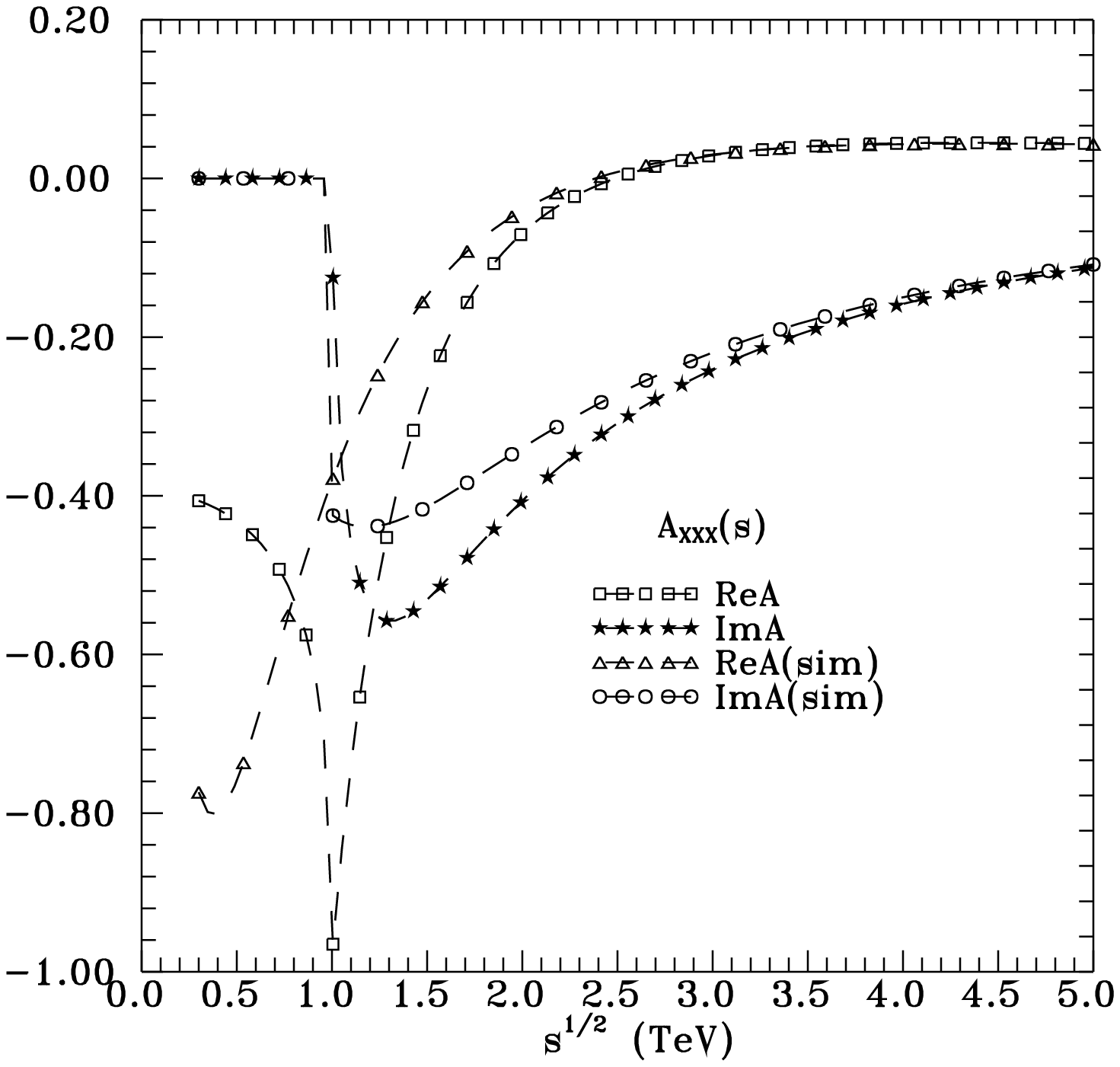,height=7.cm}
\]
\[
\epsfig{file=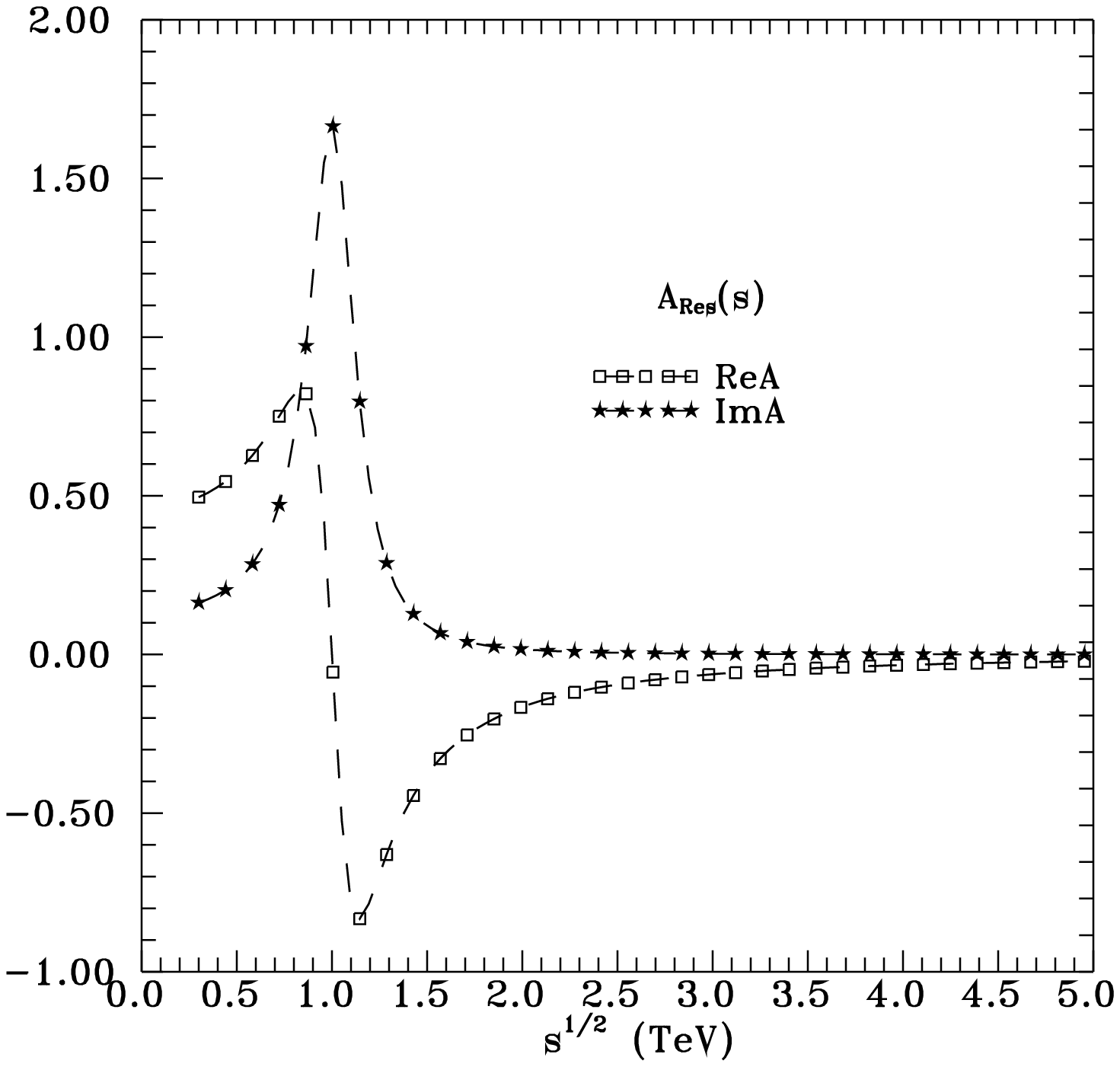, height=7.cm}
\]
\caption[1]{ New physics contributions to the  HHH form factors. Upper left  panel gives the example
of $A(s)_{FFF}$ of (\ref{ANPFFF}), upper right gives the example of $A(s)_{XXX}$ of (\ref{ANPXXX}),
while the lower panel the example of $A(s)_{\rm Res}$ of (\ref{ANPRes}). }
\label{Fig2}
\end{figure}

\end{document}